\documentclass[a4paper]{JAC2003}
\usepackage{varwidth}
\usepackage{xcolor}
\usepackage{cite}

\usepackage{graphicx}
\usepackage{booktabs}

\begin{document}
\title{BEAM--BEAM-INDUCED ORBIT EFFECTS AT LHC}

\author{M. Schaumann, RWTH Aachen, Aachen, Germany\\
R. Alemany Fernandez, CERN, Geneva, Switzerland}

\maketitle

\begin{abstract}
For high bunch intensities the long-range beam--beam interactions are strong enough to provoke effects on the orbit. As a consequence the closed orbit changes. The closed orbit of an unperturbed machine with respect to a machine where the beam--beam force becomes more and more important has been studied and the results are presented in this paper.

\end{abstract}

\section{Introduction}

In the LHC (Large Hadron Collider) the beam--beam electromagnetic force is experienced as a localized, periodic distortion when the two beams interact with each other at the collision points. This force is most important for high-brightness beams, which is the case at the LHC, and it can be classified into two types: head-on and long-range. The effects of the beam--beam force manifest themselves in very different ways. In this paper the closed-orbit effects due to long-range beam--beam interactions are studied.
Long-range interactions distort the beams much less than head-on interactions. However, there is a large number of them due to the large number of bunches per beam (2808 bunches per beam in the nominal LHC). In nominal conditions, up to 30 long-range interactions per experiment have to be expected.
Experimental data are presented in order to validate the simulation studies about the beam--beam effects at the LHC \cite{Tatiana,WernerCAS,wernerCrossingSchemes} performed during recent years. More details of the analysis presented here can be found in Refs. \cite{michaelaThesis,lr-md2}.

\section{Long-range Beam--Beam Kick}
When the beams are separated, as is the case for long-range beam--beam interactions, the beams will exert a kick to each other whose
coherent dipole component leads to orbit changes. The change in angle (kick) can be computed with the following equation \cite{Tatiana}:
\begin{eqnarray}
\Delta r' = - \frac{2N r_0}{\gamma}\cdot \frac{1}{r} \cdot \left[ 1-\exp(-\frac{r^2}{4\sigma^2}) \right],
\label{rprime}
\end{eqnarray}
where $r$ is the beam separation, $\sigma$ is the beam size at the interaction point and $\gamma$ is the relativistic Lorentz factor.

\subsection{Analysis and Results}
During a dedicated machine study period with only one bunch circulating per beam, a horizontal orbit scan in steps of 100\,$\mu$m was performed in ATLAS (IP1) and a vertical one in CMS (IP5) to measure the orbit kick due to the beam--beam force. The scan went far enough to reach the non-linear regime of the force.

\begin{figure}[htb]
   \centering
   \begin{minipage}[b]{7.5cm}
   \hspace*{-0.3cm}
   \includegraphics*[width=0.9\textwidth]{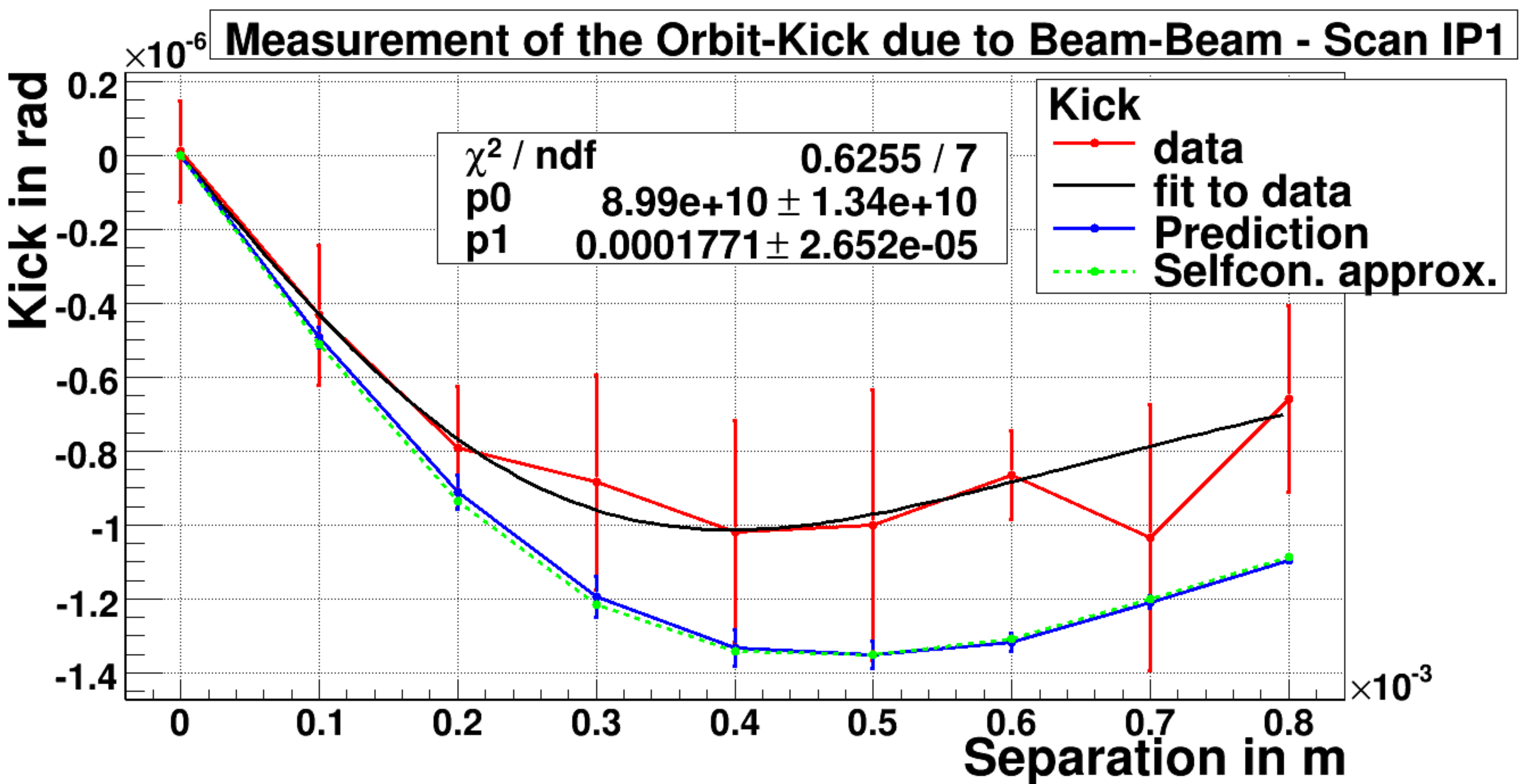}
   \end{minipage}
   \begin{minipage}[b]{7.5cm}
   \hspace*{-0.3cm}
   \includegraphics*[width=1\textwidth]{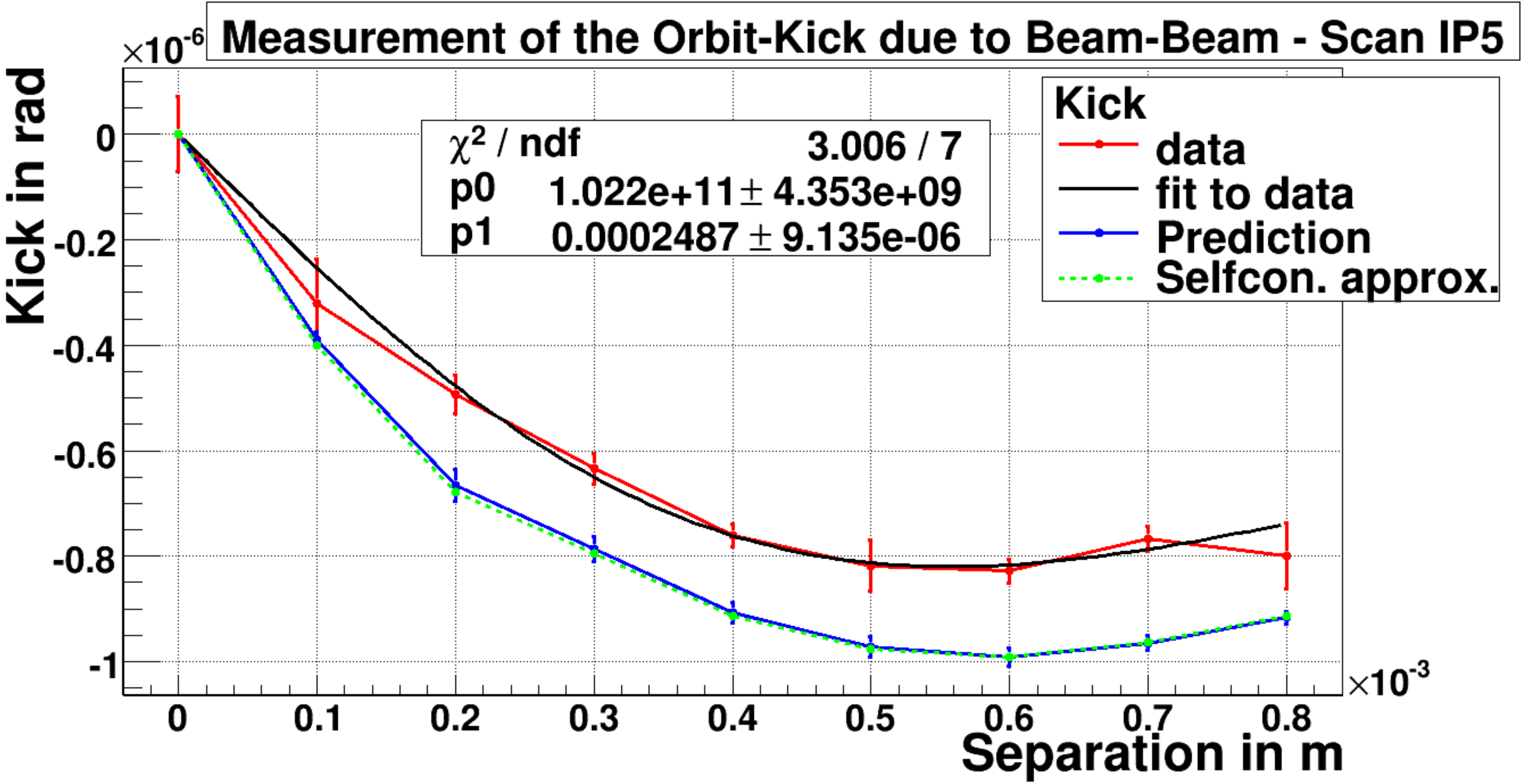}
   \end{minipage}
   \caption{Measurement and prediction of the orbit kick due to beam--beam force during an orbit scan. The vertical axis is given in units of 10$^{-6}$\,rad and the horizontal axis in units of 10$^{-3}$\,m. }
   \label{orbitkick}
   \vspace*{-0.3cm}
\end{figure}

To avoid additional effects arising from the trim of the separation bumps, only beam 1 was moved and beam 2 was untouched. In this way beam 2 is only perturbed by the beam--beam kick and a clean signal should be measured.

The result is illustrated in Fig.\,\ref{orbitkick}, where the red line indicates the kick determined by an orbit correction of the taken data and the solid blue line was calculated using Eq.\,(\ref{rprime}).

For the calculation, the values of $\sigma$ and $N$ were varied at each separation step according to the measured values. The shape of the measured curve fits to the expectation, but the measured strength is lower than the prediction. The black line is a fit to the data using Eq.\,(\ref{rprime}); the intensity $N=p_0$ and the beam size $\sigma = p_1$ were left as free but constant parameters. The obtained value for $\sigma$ is in agreement with the measurement, but the obtained $N$ is too small compared to the measured one; thus, the origin of the discrepancy must be different. Therefore, the green dashed line tries to approximate the self-consistent effect which is not considered in Eq.\,(\ref{rprime}) and arises from the effect of the interaction itself. If the orbit is changed due to a bunch crossing, the positions of the beams will be different after one turn, which implies a variation of the force, leading again to a different orbit until an equilibrium is found. The dashed line shows that this effect is too small to explain the gap. Further investigation is needed to identify the discrepancy.

\begin{figure}[ht]
   \centering
   \begin{minipage}[b]{0.49\linewidth}
   \includegraphics*[width=1\textwidth]{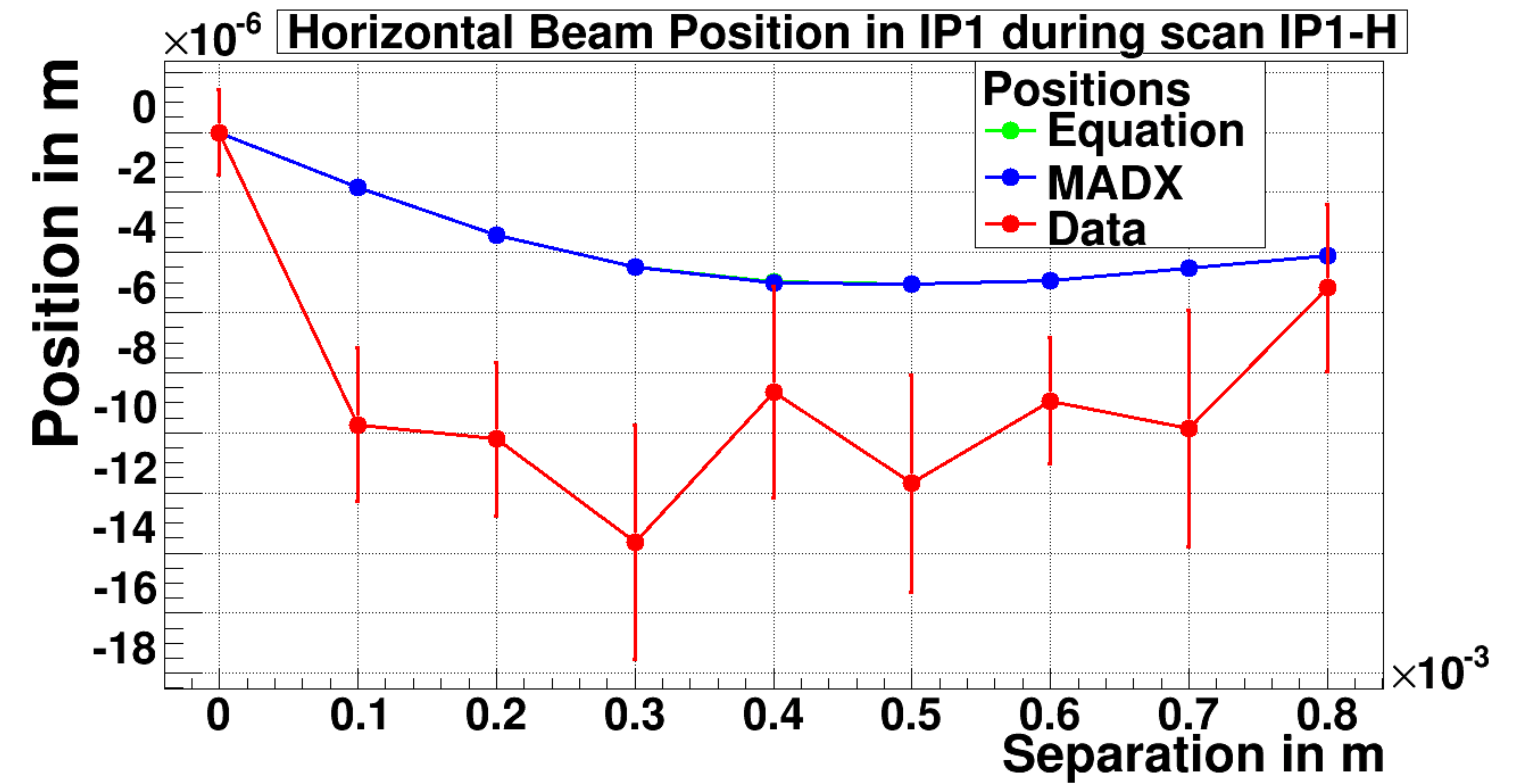}
   \end{minipage}
   \begin{minipage}[b]{0.49\linewidth}
   \includegraphics*[width=1\textwidth]{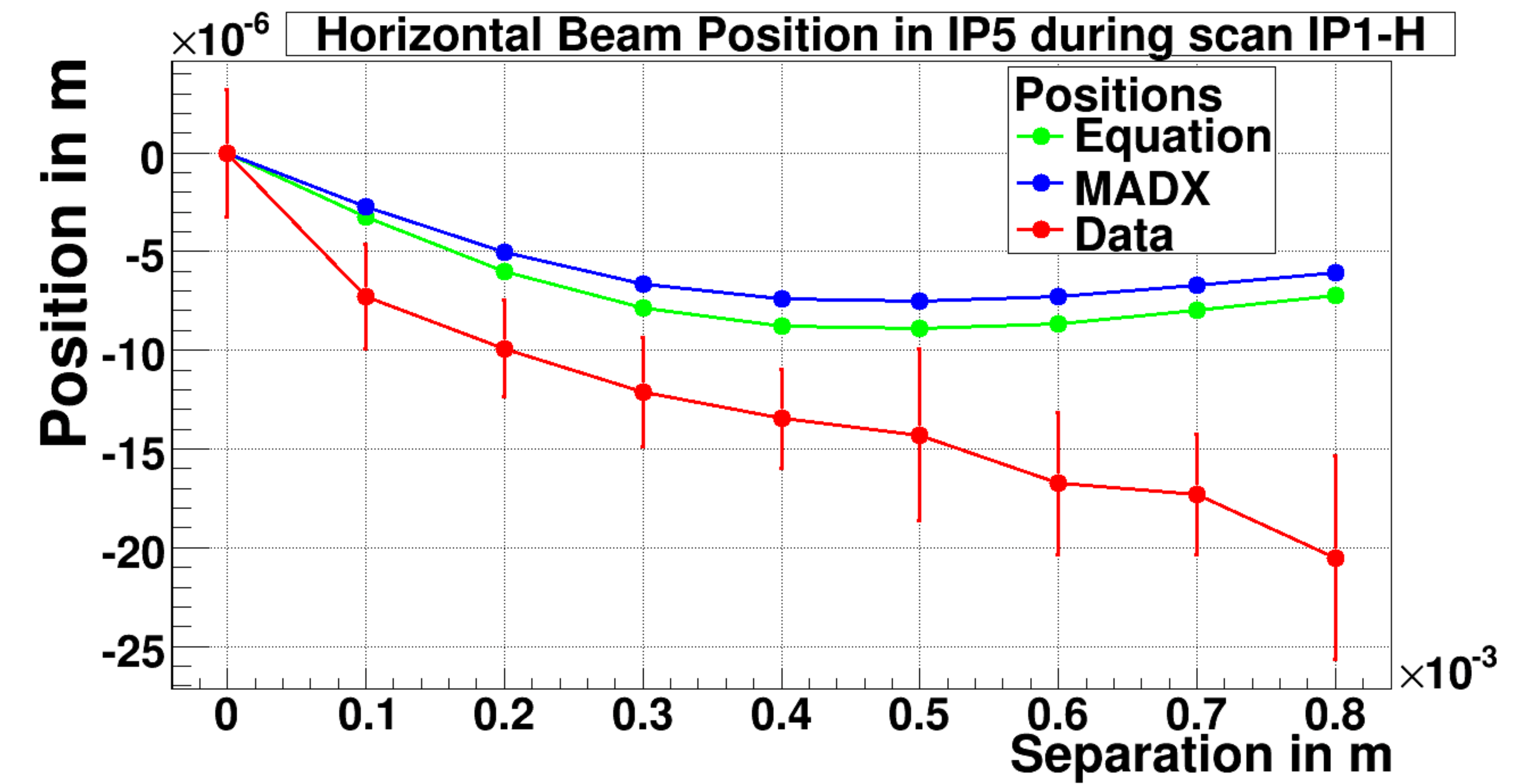}
   \end{minipage}
   \begin{minipage}[b]{0.49\linewidth}
   \includegraphics*[width=1\textwidth]{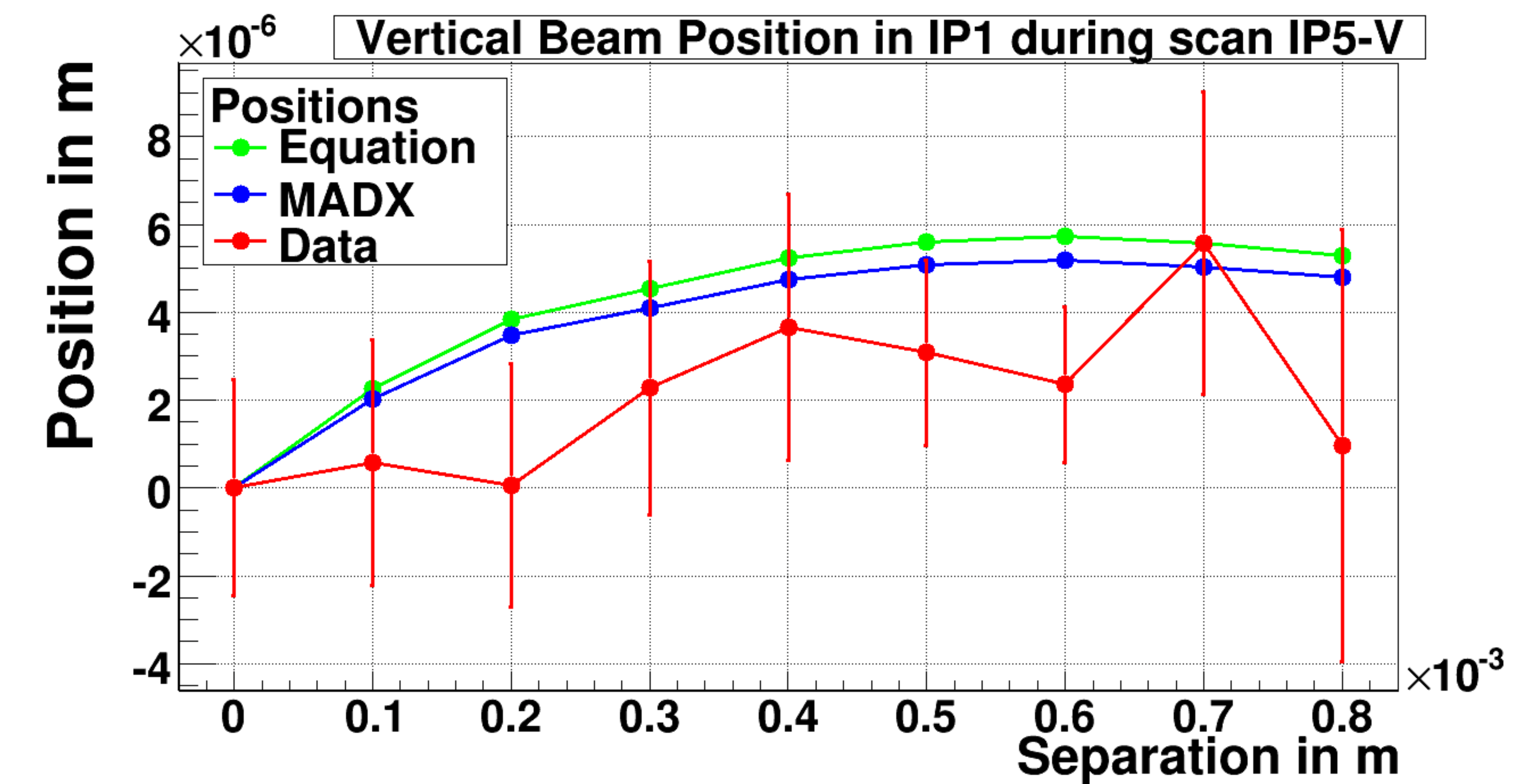}
   \end{minipage}
   \begin{minipage}[b]{0.49\linewidth}
   \includegraphics*[width=1\textwidth]{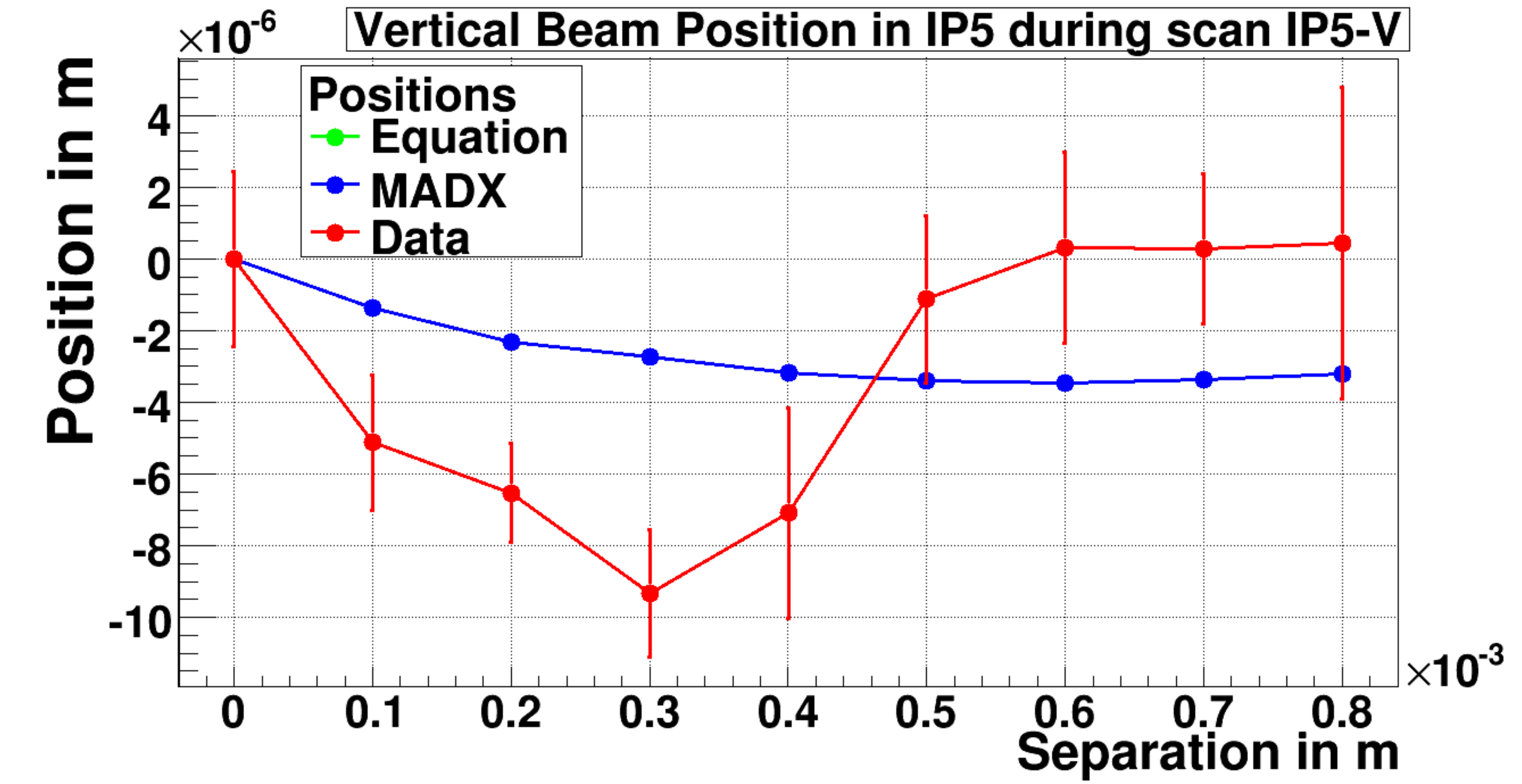}
   \end{minipage}
   \caption{Measurement and prediction of the positions at IP1 and IP5 due to beam--beam force during an orbit scan. The vertical axis is given in units of 10$^{-6}$\,m and the horizontal one in units of 10$^{-3}$\,m.}
   \label{PositionAtIp}
\end{figure}

Figure\,\ref{PositionAtIp} shows in red the beam positions at IP1 and IP5 during the scans as a function of the separation, where the error bars show the standard deviation. A prediction was calculated and simulated with the MADX program. Here a discrepancy between prediction and measurement is also found. Nevertheless, the positions at the IPs were interpolated from the data of the two closest beam-position monitors (BPMs) on the left- and right-hand sides of the IPs only. Those BPMs have a resolution of around 5\,$\mu$m, which is larger than the standard deviation and the observed discrepancy. As expected, the shape of the curves agree with the prediction (except for the beam position in IP5 during the scan of IP5, which needs further investigation). However, the source of the gap will be explained when the gap in Fig.\,\ref{orbitkick} is understood.

\section{Bunch-by-Bunch Orbit Differences}
The study of the orbit effects due to the coherent dipolar kick of the beam--beam force performed in the previous section shows that the expected orbit variations are in the order of a few micrometers. It is interesting to analyse this effect on a bunch-by-bunch basis, since the so-called PACMAN effect introduces differences bunch-to-bunch: bunches at the head or tail of a bunch train encounter fewer long-range interactions since they cross empty buckets. This leads to differences in the long-range orbit kicks for those bunches w.r.t. to the core of the train \cite{wernerCrossingSchemes}. Three ways of measuring the position variations are analysed in the following.

\subsection{ATLAS Luminous Region Reconstruction}
The analysis presented in the following uses the microvertex detector of the ATLAS experiment.
A total of five luminosity fills with different filling schemes have been analysed to investigate the dependence on the number of bunches.
 \begin{figure}[htb]
   \centering
   \begin{minipage}[b]{7.cm}
	\hspace*{-0.3cm}
   \includegraphics*[width=0.58\textwidth, angle = -90]{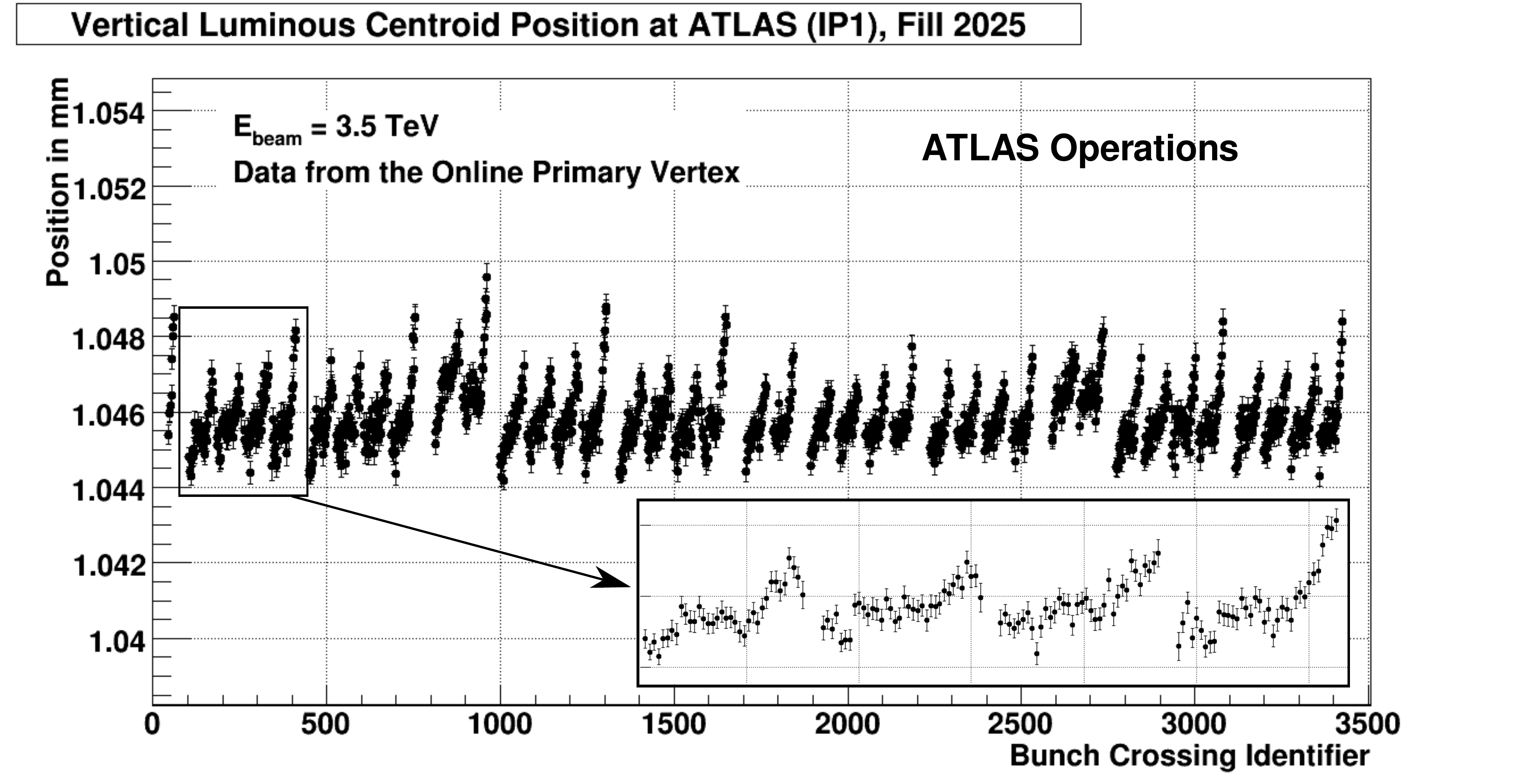}
   \end{minipage}
   \begin{minipage}[b]{7.cm}
   \hspace*{-0.3cm}
   \includegraphics*[width=0.58\textwidth, angle = -90]{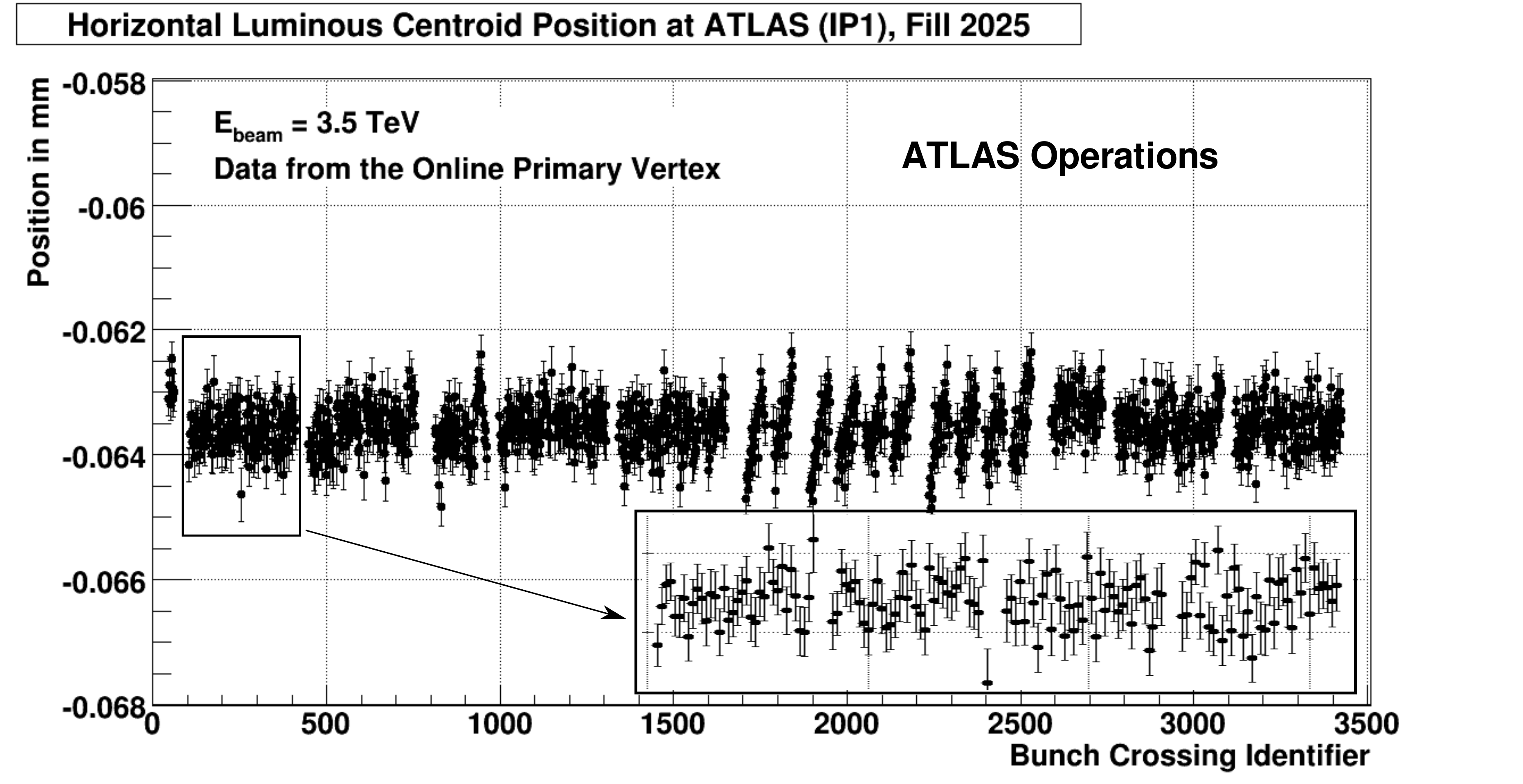}
   \end{minipage}
   \caption{Luminous centroid position in ATLAS. Top (bottom) figure shows the vertical (horizontal) luminous region position and a zoom over the first four trains of 36 bunches.}
   \label{vertexReconstruction}
\end{figure}

In Fig.\,\ref{vertexReconstruction} the reconstructed luminous region in the vertical (crossing-angle) plane and horizontal (separation) plane of ATLAS is plotted as a function of the bunch-crossing identifier for fill 2025 with 1380 bunches spaced by 50\,ns.
As predicted by the simulations \cite{wernerCrossingSchemes}, the orbits at the interaction point of all bunches are slightly different and this is more evident for PACMAN bunches in the vertical plane.

The luminous centroid reconstruction can only give the convoluted position of both beams. But, since the orbits of both beams are almost identical at the interaction point in the vertical plane (simulations \cite{wernerCrossingSchemes} show that they are slightly different due to the intensity variations bunch-by-bunch, in the order of 15\% in this fill), most of the bunches collide head-on, although not all of them in the central orbit. In particular, the PACMAN bunches show offsets w.r.t. the core of the train of (2.0 $\pm$ 0.3)\,$\mu$m due to the different number of long-range interactions. There are two types of them, the ones close to an eight-bunch gap (between trains), and the ones close to a 36/56-bunch gap (between groups of two or four trains). In Fig.\,\ref{vertexReconstruction} a zoom over the first four 36-bunch trains is illustrated. The bunches at the core of the train show a different structure than the bunches close to a gap. This is due to the different number of long-range interactions. While the bunches at the core of the train have on average 74 long-range interactions, the PACMAN bunches have in the order of 10 long-range interactions fewer. The different number of long-range interactions implies a different amount of coherent dipolar kicks, and therefore different orbits. The effect is even more clear for the PACMAN bunches close to the 36/56-bunch gap; those have on average 40 long-range interactions fewer and the orbit difference is, consequently, bigger.

The orbit offset due to beam--beam kicks in the vertical plane is (5.0 $\pm$ 0.3)\,$\mu$m peak to peak. This is a factor of 2.5 bigger than the offsets presented in the simulations of Ref. \cite{wernerCrossingSchemes}. Nevertheless, it has to be taken into account that the simulations are done for nominal LHC parameters, i.e. 2808 bunches spaced by 25\,ns; therefore, the bunch pattern is different. As a continuation of this study, new simulations will be done with the parameters of this fill to be able to assess the data in a more quantitative way.

The orbit offsets in the horizontal plane of IP1 do not show a particular structure, as in the case of the vertical plane. If the orbit could be reconstructed for both beams separately, simulations predict that the positions of the beams are symmetric to the central orbit and offset by a few micrometers (see Ref. \cite{wernerCrossingSchemes}, Figs. 14--16). The PACMAN bunch effect appears as well, but in the horizontal plane it arises from the long-range interactions in the vertical (crossing) plane of IP5. Since the effect in this plane is global and travels through the ring, it appears with a different shape as the local effect in the crossing-angle plane of the considered IP.
But, since we are using the luminous region, only the convolution of both beams is visible, and the effect is cancelled because of its symmetry. There are, however, some trains which show a structure compatible with the beam--beam effect of the vertical plane. The source of this structure should come from extra coherent dipole kicks happening in the horizontal plane somewhere else for those bunches, but this effect needs further investigation.

No dependence on the number of bunches was found for the filling schemes analysed with the same number of long-range interactions, as could be expected.

\subsection{Luminosity Optimization}
\begin{figure}[htb]
   \centering
   \begin{minipage}[b]{7cm}
   \hspace{-0.3cm}
   \includegraphics*[width=1.2\textwidth, angle = 0]{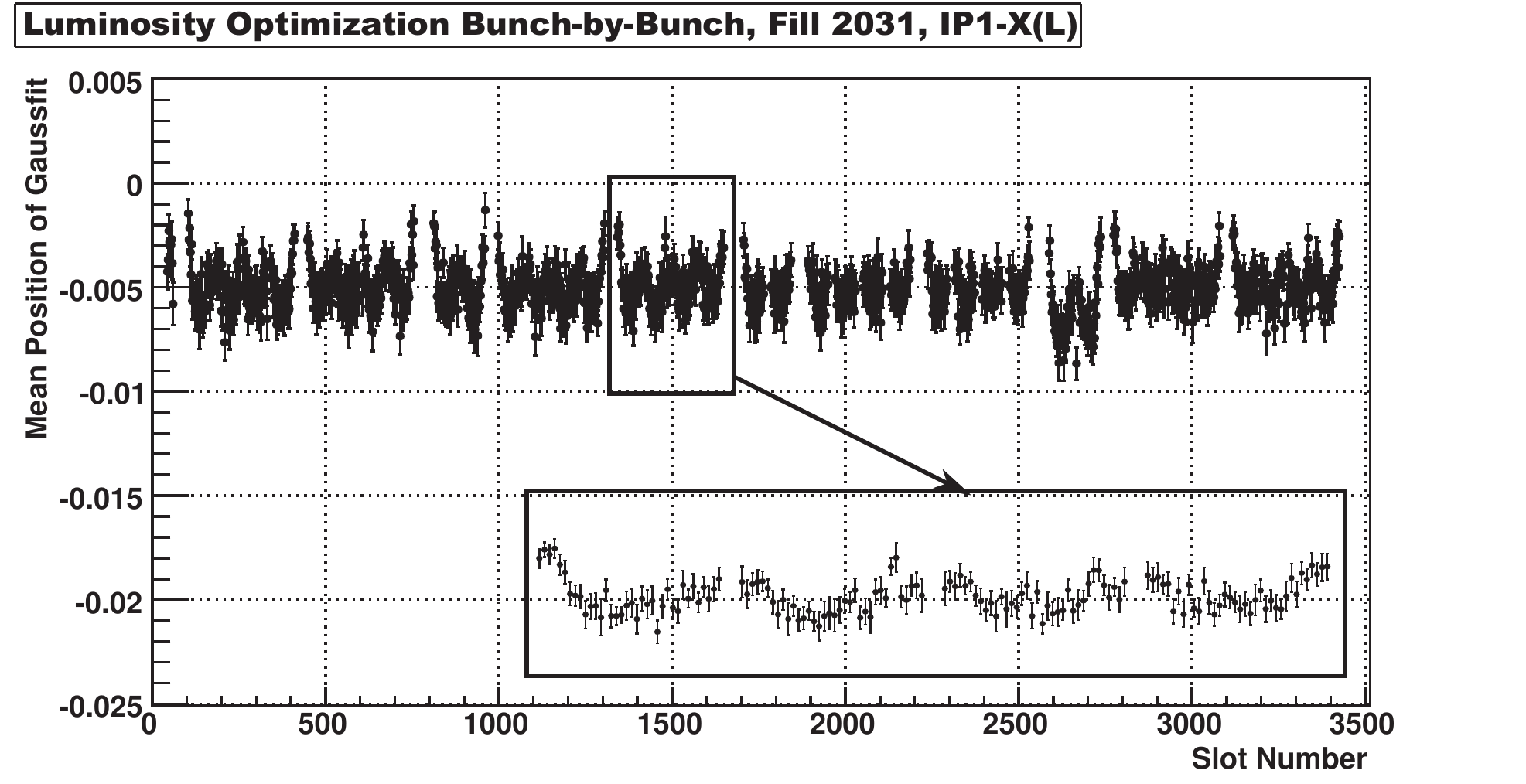}
   \end{minipage}
   \begin{minipage}[b]{7cm}
   \hspace{-0.3cm}
   \includegraphics*[width=1.2\textwidth, angle = 0]{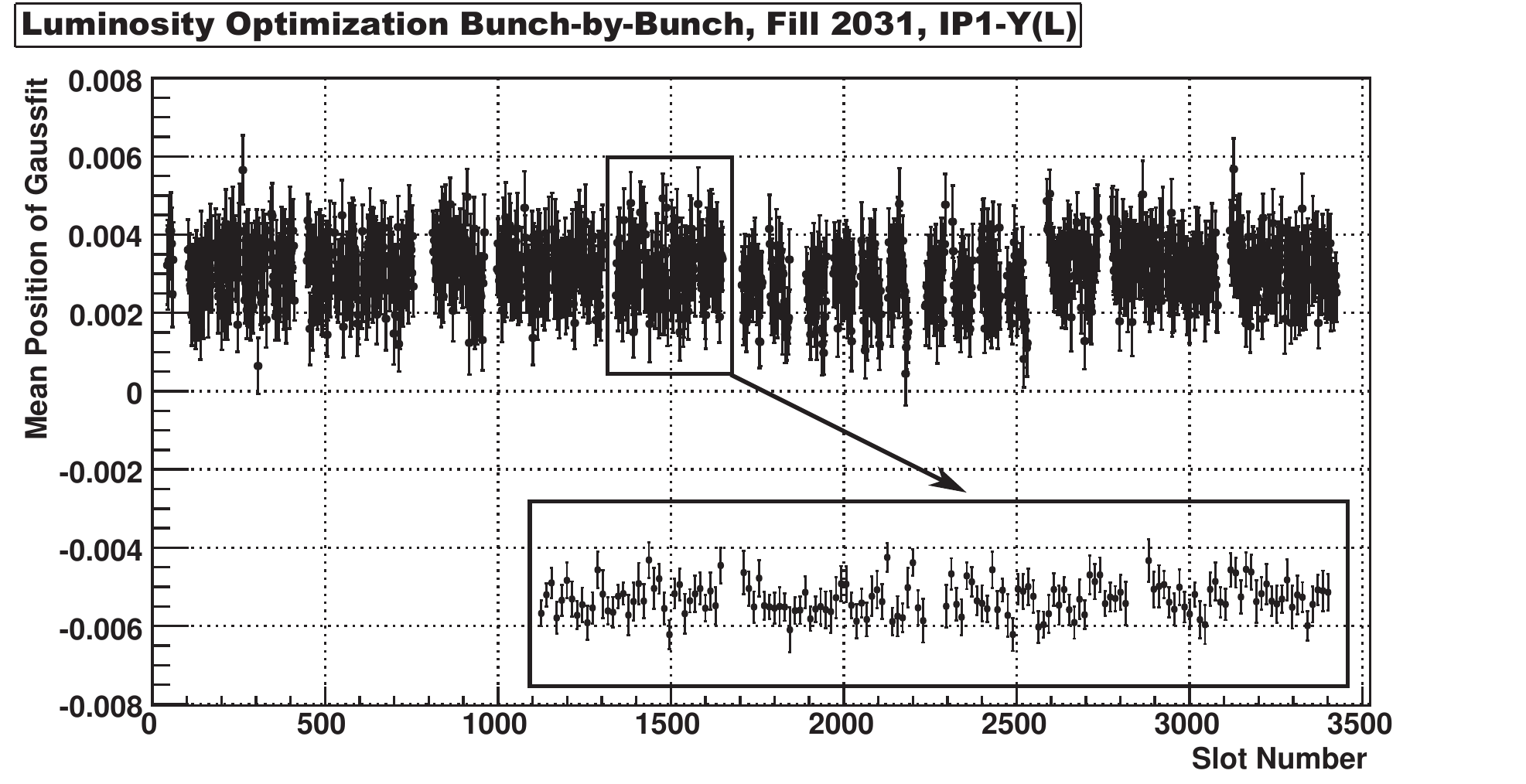}
   \end{minipage}
   \caption{Position of the maximal luminosity (total bunch overlap) as a function of the bunch number. Top (bottom) figure shows the horizontal (vertical) mean position of the luminosity maximum in units of millimeters w.r.t. the beam position before the scan.}
   \label{lumiScan}
\end{figure}

During every luminosity run, once the beams are brought into collision, every interaction point undergoes a luminosity scan in the horizontal and vertical planes to find the full beam overlap and therefore the maximum luminosity. This process is done by integrating the luminosity over the whole beam, and assuming an average position over all bunches. However, since we  have demonstrated that there are bunch-by-bunch orbit differences, it is interesting to look at the variation in the maximum of the luminosity as a function of the bunch number. Figure\,\ref{lumiScan} displays the relative position of the two colliding bunches where the maximum luminosity is found during the scan. The plots show IP1 in the horizontal plane (top plot) and vertical plane (bottom plot).

To understand this structure, the position of each beam should be reconstructed. Since this is not possible with the BPMs, simulations have to be referenced again (see Ref. \cite{wernerCrossingSchemes}, Figs. 14--16): the PACMAN bunch positions are offset w.r.t. the core of the train and have a mirror-like structure to the opposite beam. When the scan is performed, the luminosity maximum is first found for those bunches and afterwards for the core of the train. Thus, in Fig. \ref{lumiScan} (top) the Gaussian fit mean position is closer to the zero displacement for the PACMAN bunches.

No structure is visible in the vertical plane, because in IP5 the vertical plane is the separation plane where there are no long-range effects. Moreover, the local long-range effects in IP1 cannot be seen as in Fig.\,~\ref{vertexReconstruction}, since as explained above the orbits for both beams are nearly identical and as both beams are moved symmetrically in opposite directions the effect is not visible using this measurement method, since the separation is the same for all bunches.

\subsection{Beam-position Monitors}
An exhaustive analysis of the data measured by the LHC BPM system during a typical luminosity run was performed to assess the feasibility of those instruments to measure closed-orbit deviations in bunch-by-bunch mode in the order of a few micrometers.

From the analysis one can conclude that the BPM system has a bunch-by-bunch orbit measurement able to give a relative bunch-to-bunch resolution in the 5\,$\mu$m range. Yet the non-linearity in the bunch-by-bunch mode for different global positions limits this resolution to $\pm$50\,$\mu$m when comparing along the train for different mean positions.
The current LHC BPM system, therefore, does not have sufficient linearity or resolution to resolve the bunch-by-bunch orbit variations at the few-micrometer level expected from beam--beam interaction orbit effects during normal operation.

However, if the separation at the parasitic encounters is decreased, the long-range beam--beam force is enhanced and the orbit distortion becomes strong enough to be measured with the BPM system. In a dedicated experiment \cite{lr-md,lr-md2} the crossing angle was reduced simultaneously in IP1 and IP5 by the same amount and bunch-by-bunch orbit data were taken during every step of the experiment.

The bunch-by-bunch orbit measurement is the sum of many components, e.g. common motions of all bunches, like betatron oscillation and initial orbit differences between bunches; moreover, electronic and temperature effects influence the absolute position measured between bunches and BPMs. In this experiment the interest lies in the observation of the orbit changes introduced by the enhanced long-range kick and not in the absolute position of the beam. Therefore, all those effects influencing the absolute beam position have to be filtered out to resolve the small effects of the changing beam--beam force from the BPM measurement.

In the following, only the data of beam 1, consisting of three trains with 12, 36 and 36 bunches, are plotted. The 12 bunches of train 1 were all non-colliding and train 3 only collides in IP8. Train 2 collides in IP1 and IP5, where the crossing angle was varied, and additionally in IP2. Bunches in train 2 had up to 16 long-range interactions per IP.

\begin{figure}[htb]
   \centering
   \begin{minipage}[b]{8cm}
   \includegraphics*[width=0.6\textwidth, angle = 90]{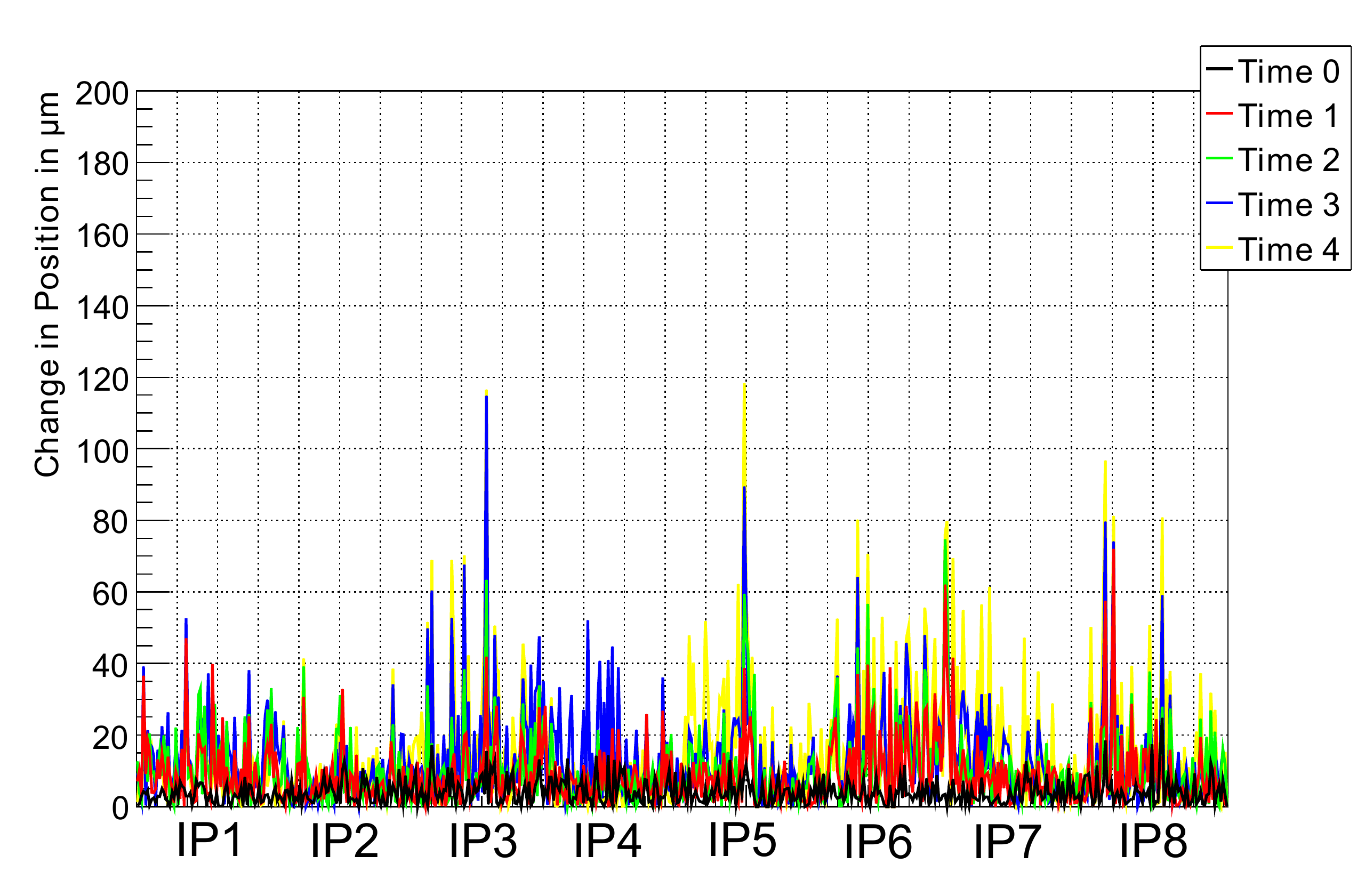}
   \vspace{-0.2cm}
   \end{minipage}

   \begin{minipage}[b]{8cm}
   \includegraphics*[width=0.6\textwidth, angle = 90]{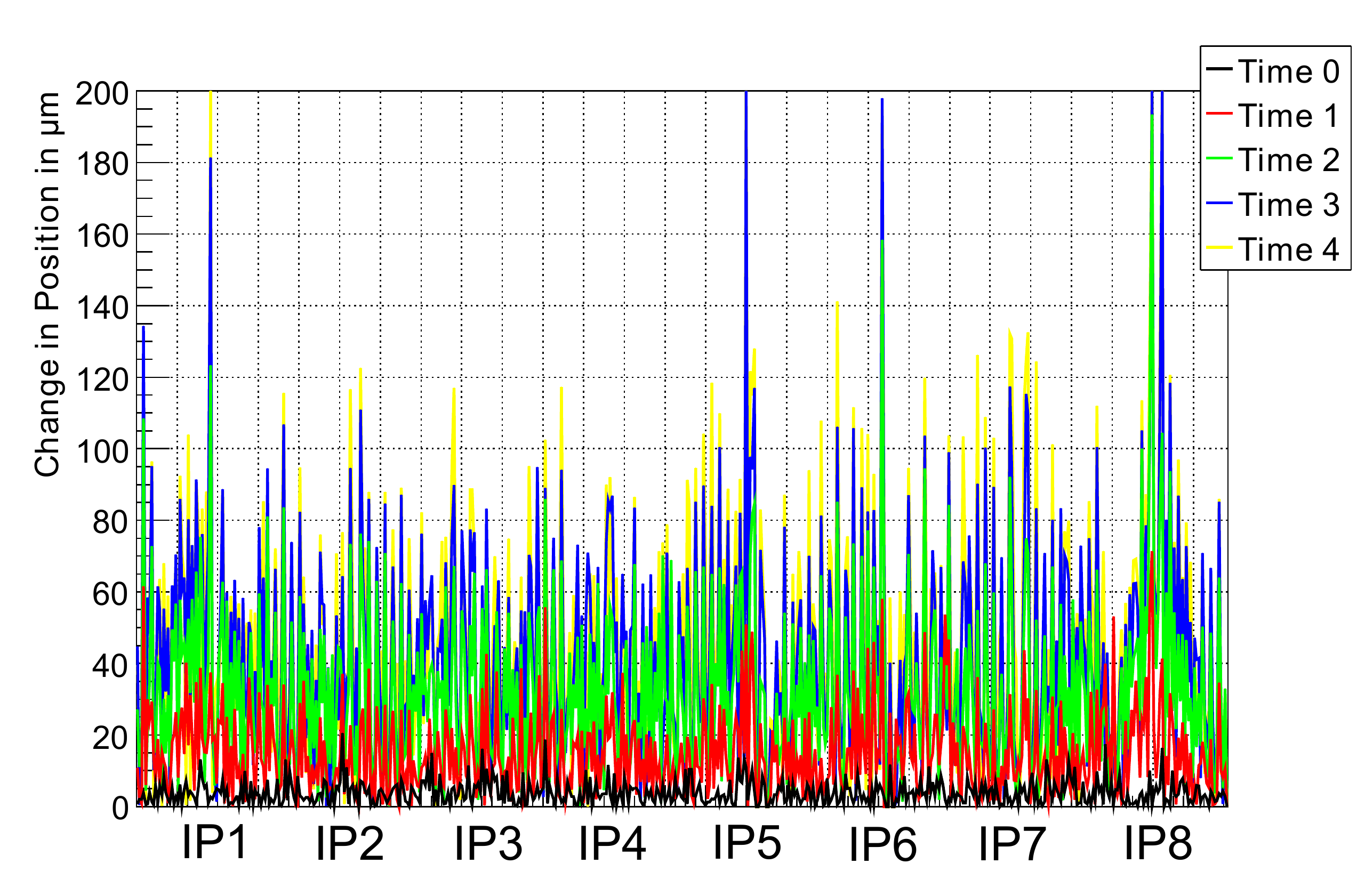}
   \end{minipage}
   \caption{Horizontal absolute change in position of beam 1 as a function of the location in the ring for five different crossing angles. Top: bunch with zero head-on and long-range interactions, bottom: three head-on, 16 long-range interactions.}
   \label{orbitRing}
\end{figure}

Figure\,~\ref{orbitRing} shows the changes in relative position (after filtering) for different crossing angles (times 0 to 4) as a function of the location around the ring. In the top plot a non-colliding bunch (with zero head-on and zero long-range interactions) is shown with respect to a reference bunch with the same properties. Since this bunch does not experience any long-range interactions, it is not expected to change its position when the crossing angle is decreased from times 0 to 4. A noise floor of $\propto 20$--40\,$\mu$m is clearly visible, which limits the resolution. This is in agreement with the value found during normal operation.

The bottom plot shows a bunch of train 2 with three head-on and 16 long-range interactions with respect to the same reference bunch. In this case a clear systematic structure above the noise level develops which increases when the crossing angle is decreased, giving a qualitative measurement of the enhancement of the long-range beam--beam force.

\begin{figure}[htb]
   \centering
   \hspace{-0.2cm}
   \includegraphics*[width=0.45\textwidth, angle = 0]{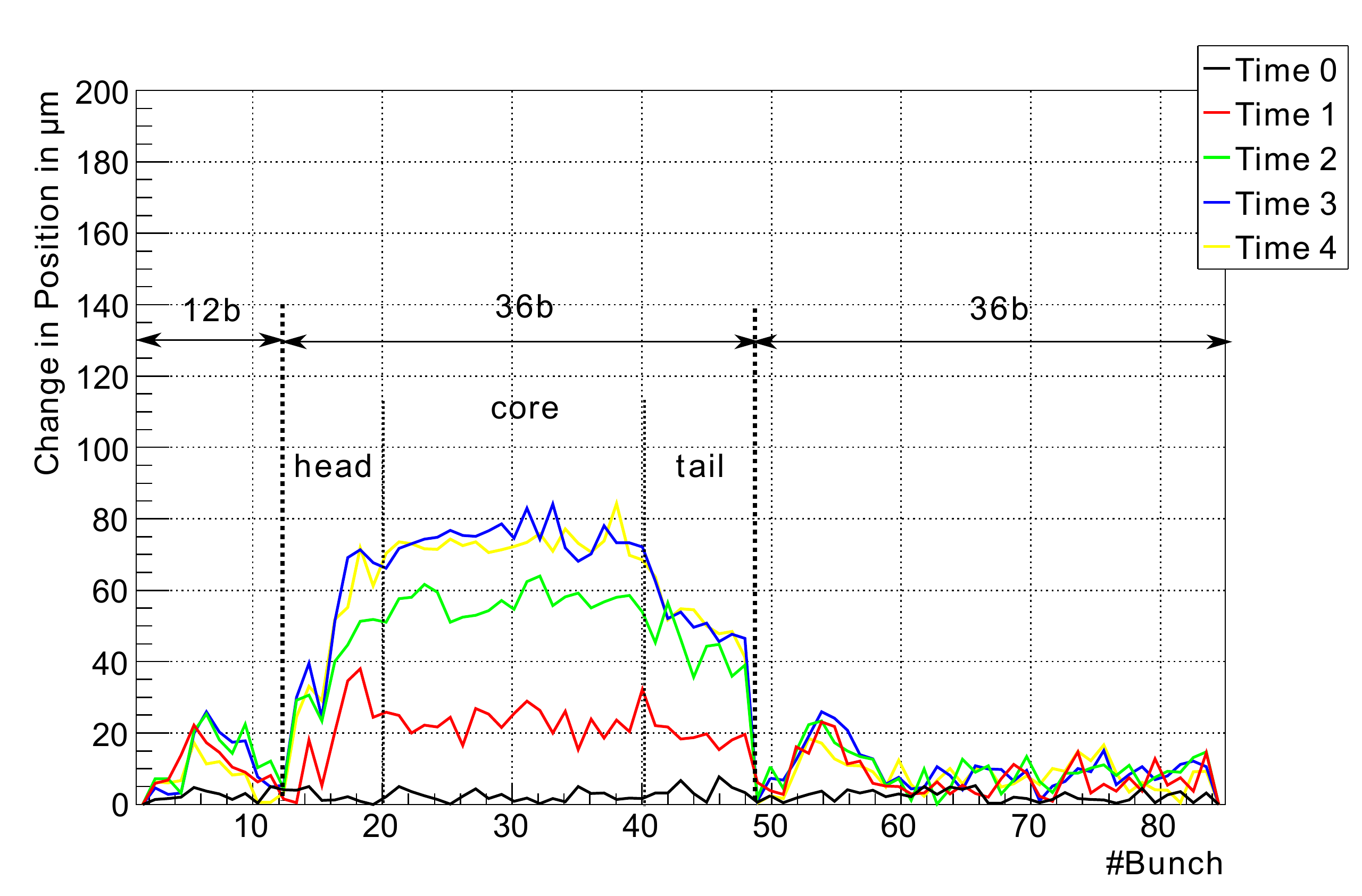}
   \caption{Horizontal absolute change in position of beam 1 as a function of the bunch number for a particular BPM around IP1.}
   \label{orbitBunch}
\end{figure}

In Fig.\,~\ref{orbitBunch} the relative position change at varying crossing angles is shown at a particular BPM (BPM.6L1.B1) on the left-hand side of IP1 for all bunches in the machine. Train 2 shows the typical PACMAN bunch behaviour: bunches in the core of the train experience the largest number of long-range kicks and therefore show the largest change in position while reducing the crossing angle. With decreasing number of long-range interactions to the ends of the train, the effect on the orbit is also reduced. Since only the crossing angles in IP1 and IP5 were varied, trains 1 and 3, which do not have collisions in those IPs, are not affected.

\section{Acknowledgements}
The authors would like to thank the LHC Operations Group and the Beam--Beam Working Group for their support during the machine development studies and J. Wenniger, M. Favier, R. Jones, E. Calvo, T. B\"ar, G. M\"uller, W. Herr and B. Holzer for their fruitful discussions during the analysis of the data. We thank the ATLAS Collaboration for providing the beam-spot information from their data. In particular, we are grateful to R. Bartoldus, J. Cogan, C. Gwilliam, E. Strauss and C. Rembser for their help.


\begin{thebibliography}{9}   

\bibitem{Tatiana} T. Pieloni,  CERN-THESIS-2010-056 (2008).

\bibitem{WernerCAS} W. Herr,
CERN-2006-002 (2006), pp. 379ff.

\bibitem{wernerCrossingSchemes} W. Herr,
LHC Project Report 628 (2003).

\bibitem{michaelaThesis} M. Schaumann, CERN-THESIS-2011-138 (2011).

\bibitem{lr-md2} M. Schaumann {et al.},
CERN-ATS-Note-2012-021 MD (2012).

\bibitem{lr-md} R. Alemany {et al.},
CERN-ATS-Note-2011-120 MD (2011).

\bibitem{michaelaIpac} M. Schaumann, IPAC12, WEPC081 (2011).

\end{thebibliography}
\end{document}